\documentclass[12pt]{iopart}
\usepackage{setstack}
\usepackage{graphicx,amssymb,amsthm,epsfig,amsbsy,amsfonts,mathrsfs}
\usepackage[english,francais]{babel}

\newfont{\ensmathquatorze}{msbm10 scaled 1400}
\newfont{\ensmathonze}{msbm10 scaled 1100}
\newfont{\ensmathdix}{msbm10}
\newfont{\ensmathneuf}{msbm10 scaled 833}
\newfont{\ensmathhuit}{msbm10 scaled 694}
\newfam\ensmathfam                        
\textfont\ensmathfam=\ensmathonze        
\scriptfont\ensmathfam=\ensmathdix       
\scriptscriptfont\ensmathfam=\ensmathhuit

\def\eqdef{\stackrel{\mbox{\tiny def}}{=}}     
\newcommand{\ket}[1]{|\kern.3ex#1\kern.3ex\rangle}
\newcommand{\bra}[1]{\langle\kern.3ex #1 \kern.3ex|}
\newcommand{\mean}[1]{\overline{#1}}
\newcommand{\smean}[1]{\overline{#1}}
\newcommand{\expect}[1]{\mathbb{E}\left(#1\right)}

\newcommand{\EXP}[1]{e^{#1}}         

\def\I{{\rm i}}                  
\def\D{{\rm d}}                  

\newcommand{\deriv}[2]{\frac{\mathrm{d}#1}{\mathrm{d}#2}}

\newcommand{\diagram}[3]{\raisebox{#3}{\includegraphics[scale=#2]{#1}}}

\begin{document}

\renewcommand{\labelitemi}{$\bullet$}
\renewcommand{\labelitemii}{$\star$}

\selectlanguage{english}

\title[Ordered spectral statistics for 1D random SUSY Hamiltonian]{Ordered spectral statistics in 1D disordered supersymmetric quantum mechanics and Sinai diffusion with dilute absorbers}



\author{Christophe Texier}

\address{Univ. Paris Sud ; CNRS ; LPTMS, UMR 8626 \& LPS, UMR 8502 ; 
             Orsay F-91405, France}

\eads{ \mailto{christophe.texier@u-psud.fr}}
\begin{abstract}
  Some results on the ordered statistics of eigenvalues for
  one-dimensional random Schr\"odinger Hamiltonians are reviewed.
  In the case of supersymmetric quantum mechanics with disorder,
  the existence of low energy delocalized states induces
  eigenvalue correlations and makes the ordered statistics problem
  nontrivial. 
The resulting distributions are used to analyze the problem
  of classical diffusion in a random force field (Sinai problem) in
  the presence of weakly concentrated absorbers. It is shown that the
  slowly decaying averaged return probability of the Sinai problem,
  $\mean{P(x,t|x,0)}\sim \ln^{-2}t$, is converted into a power law
  decay, $\mean{P(x,t|x,0)}\sim t^{-\sqrt{2\rho/g}}$, where $g$ is the
  strength of the random force field and $\rho$ the density of
  absorbers.
\end{abstract}




\maketitle

\section{Introduction}

An ordered (or extreme value) statistics problem can be defined as
follows~: given a set of $\mathcal{N}$ ranked random variables
$x_1<x_2<\cdots<x_n<\cdots<x_\mathcal{N}$, one asks for the probability 
density $W_{n,\mathcal{N}}(x)$ for finding $x_n$ at $x$.
When the random variables are independent and identically distributed
(i.i.d) according to a given distribution $p(x)$,
the distributions $W_{n,\mathcal{N}}(x)$ may be easily related to $p(x)$. 
The interesting point emerges when considering the $\mathcal{N}\to\infty$
limit~: in this case, up to a rescalling
$x_n=a_{n,\mathcal{N}}+b_{n,\mathcal{N}}\,y_n$, 
that depends on the original distribution $p(x)$, the distribution of
the rescaled variable $y_n$ converges to an {\it universal} law $w_n(y)$. 
Three universality classes exist, corresponding to the nature of the
tail of $p(x)$, as demonstrated in the pioneering works of
Fr\'echet \cite{Fre27} and Gumbel (see the textbook \cite{Gum58}). 
When $p(x)$ has a tail of exponential type, one obtains the famous
generalized Gumbel laws $w_n(y)=\frac{n^n}{(n-1)!}\,\exp n(y-\EXP{y})$.
The study of extreme events is relevant in many areas like
meteorology, finance, computer science, statistical physics, etc, and
in general the random variables of interest may not fulfill the
hypothesis of statistical independence (see for example
Ref.~\cite{MajKra03} or \cite{BirBouPot07}).  
In this case the problem is much more difficult to analyze and no
general results are known such as for i.i.d random variables. 
This situation occurs when considering the set of eigenvalues of a
random operator, since eigenvalues are {\it a priori} correlated. 
For example, the distribution of the smallest eigenvalue of large
random matrices 
has been obtained in a famous work by Tracy and Widom
\cite{TraWid94,TraWid96} 
for the usual Gaussian ensembles~;
other ensembles have been recently considered in
Refs.~\cite{Joh00,Joh01,VivMajBoh07,NadMajVer11}. 
In the present paper, we will consider the case where random variables
are eigenvalues of a one-dimensional (1D) Schr\"odinger operator with 
a random potential.

The paper is organised as follows~: in section \ref{sec:oss} we define
the problem and review some results. In particular we will focus
ourselves on the case of disordered {\it supersymmetric} quantum
mechanics for 
which we obtain a set of non trivial distributions $w_n(y)$. 
Applications of these results are discussed in sections
\ref{sec:breaksusy} and \ref{sec:sinaibs}~: 
in the context of 1D disordered quantum mechanics and in the context
of classical 
diffusion in a random force field with absorption.



\section{Ordered spectral statistics}
\label{sec:oss}

Let us consider the Schr\"odinger operator
$H_\mathrm{scalar}=-\deriv{^2}{x^2}+V(x)$ 
acting on functions defined on the interval $[0,L]$, satisfying
Dirichlet boundary conditions $\varphi(0)=\varphi(L)=0$. We denote by 
$E_1<E_2<\cdots<E_n<\cdots$ the (infinite) set or eigenvalues.
We will be interested in situations where the potential $V(x)$ is
random. The section is devoted to the analysis of the probability
density for finding the $n$-th eigenvalue $E_n$ at level $E$, denoted
$W_n(E;L)=\mean{\delta(E-E_n)}$, 
where $\mean{\cdots}$ denotes averaging with respect with the disordered
potential $V(x)$.
The knowledge of this set of distributions provides some
spectral informations much more precise than the averaged density of
states (DoS) $\mean{\varrho(E;L)}=\sum_{n=1}^\infty W_{n}(E;L)$, 
measuring the probability to find {\it any} eigenvalue at level $E$.
The determination of the distributions $W_n(E;L)$ has been achieved
for different types of random potentials, what we review in the rest
of the section.
We will consider the limit $L\to\infty$ (in practice $L$ must
be larger than a certain scale charcateristic of the disorder) 
analogous to the $\mathcal{N}\to\infty$ limit of the introduction.

\subsection{Fully localized spectrum}

The distributions $W_n(E;L)$ were determined when $V(x)$ is a
Gaussian white 
noise~: the distribution of the ground state energy ($n=1$) was obtained in
Ref.~\cite{McK94}, a result generalized to all eigenvalues in
Ref.~\cite{Tex00}. 
The case of a potential describing the superposition of repulsive
scatterers at random positions, $V(x)=\sum_i\alpha_i\,\delta(x-x_i)$,
was considered in Ref.~\cite{GreMolSud83}~;
$x_i$ are uncorrelated and uniformly distributed on $[0,L]$ with
a mean density $\rho$~; the positive weights $\alpha_i$ can be
chosen random or not.
These two models of random potential present different spectral
densities and localization properties. They share however two
facts~\cite{LifGrePas88}~: 
({\it i}) a suppression of the low energy DoS (Lifshits
tails), compared to the free DoS, and 
({\it ii}) a decreasing localization length as $E$ decreases (low
energy states get more localized).  
This explains the origin of the result obtained in
Refs.~\cite{GreMolSud83,Tex00}~: after an appropriate model dependent
rescaling of energies, $y_n=(E_n-\mean{E_n})/\sqrt{\mathrm{var}(E_n)}$, 
the distributions $W_n(E;L)$ coincide, in the limit $L\to\infty$,
with the generalized Gumbel laws $w_n(y)$ aforementionned for {\it
  uncorrelated} 
variables. The absence of energy level correlations is explained by the
strong localization of low energy states trapped in deep potential
wells~\cite{Mol81}.

\subsection{Spectrum unfolding}

An interesting observation is related to the structure obtained in
Ref.~\cite{Tex00} for localized eigenstates in the
$L\to\infty$ limit~:
$W_n(E;L)\simeq LN'(E)\frac1{(n-1)!}[LN(E)]^{n-1}\exp-LN(E)$, where
$N(E)=\lim_{L\to\infty}\frac1L\int^E_{-\infty}\D{}E'\,\mean{\varrho(E';L)}$ is 
the integrated DoS per unit length (IDoSpul) for
$L=\infty$. The change of variable $\xi_n=LN(E_n)$ corresponds to
unfolding the spectrum~: it relates a set of random variables $\{E_n\}$
distributed with a 
non-uniform density $\mean{\varrho(E;L)}\simeq LN'(E)$, to random
variables $\{\xi_n\}$ distributed with a uniform density equal to unity. 
The fact that these latter variables are distributed according to a
Poisson law $\varpi_n(\xi)=\frac1{(n-1)!}\xi^{n-1}\EXP{-\xi}$
demonstrates the absence of energy level correlations~\cite{Mol81}, 
since $\varpi_n(\xi)$ is also the probability that $n-1$ eigenvalues
lie in the interval $[0,\xi]$.

\subsection{Supersymmetric case}

The results reviewed in the previous subsection apply to the generic situation where 
all eigenstates are localised. However this does not exhaust all possible scenarii, 
as there exist one-dimensional disordered models exhibiting delocalized states.
Such an example is provided by the supersymmetric quantum Hamiltonian 
\begin{equation}
  \label{eq:Hsusy}
  H_\mathrm{susy} = -\deriv{^2}{x^2} + \phi(x)^2 + \phi'(x) = Q^\dagger Q
\end{equation}
where $Q=-\deriv{}{x} + \phi(x)$. 
This Hamiltonian appears in many interesting contexts
(see Refs.~\cite{ComTex98,ComDesTex05,TexHag10} for reviews). 
When $\phi(x)$ has short range correlations and
$\mean{\phi(x)}=0$, we obtain low energy properties opposite to the
one mentioned previously for $H_\mathrm{scalar}$~: 
({\it i}) the disorder {\it increases} the low energy DoS (Dyson singularity)
and  ({\it ii}) the states get {\it delocalized} near the spectrum
boundary at $E=0$ (this was demonstrated in Ref.~\cite{BouComGeoLeD90}
when 
$\phi(x)$ is a Gaussian white noise). 
The delocalization of the first eigenstates are responsible for energy
level correlations and the distributions $W_n(E;L)$ are no longer expected
to coincide with the generalized Gumbel laws.
We now explain the principle of the method, introduced in
Ref.~\cite{Tex00}, 
allowing for the determination of these distributions, when $\phi(x)$
is a Gaussian white noise such that  $\mean{\phi(x)}=0$
and $\mean{\phi(x)\phi(x')}=g\,\delta(x-x')$.

The starting point consists in converting the Sturm-Liouville
(spectral) problem, $H_\mathrm{susy}\varphi(x)=E\varphi(x)$ with boundary
conditions $\varphi(0)=\varphi(L)=0$, into a Cauchy problem, 
$H_\mathrm{susy}\psi(x;E)=E\psi(x;E)$ with boundary
conditions $\psi(0;E)=0$ and $\psi'(0;E)=1$. The former admits
solutions only for energy in a discrete set $E\in\{E_n\}$, while the
latter has solutions $\forall E\in\mathbb{R}$.
We introduce the notation $\ell_1$, $\ell_2$, etc, for the lengths
between the consecutive nodes of the wave function (Fig.~\ref{fig:psi1to3}), and $P(\ell)$
their common distribution, that will be obtained by studying the
statistical properties of $\psi(x;E)$, or rather of some related
Riccati variable. Then the probability for the eigenenergy $E_n$ to be
at level $E$ coincides with the probability for the $n$-th zero to
coincide with the boundary $\sum_{m=1}^n\ell_m=L$, i.e.
$W_n(E;L)\propto(P*\cdots*P)(L)$, as illustrated on Fig.~\ref{fig:psi1to3}.
This is the essence of the node counting method, also denoted
phase formalism in the context of disordered systems
\cite{LifGrePas88} (or Dyson-Schmidt method for discrete
models~\cite{Luc92}). 

\def\scale{0.7}
\begin{figure}[!ht]
  \centering
  $E<E_1$~: \diagram{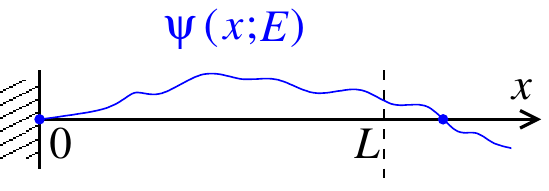}{\scale}{-0.35cm}\\
  $E=E_1$~: \diagram{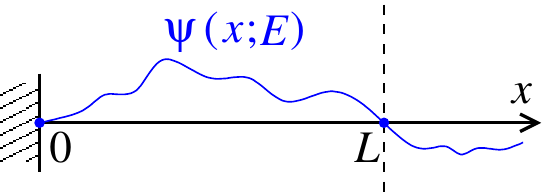}{\scale}{-0.35cm}\\
  $E>E_1$~: \diagram{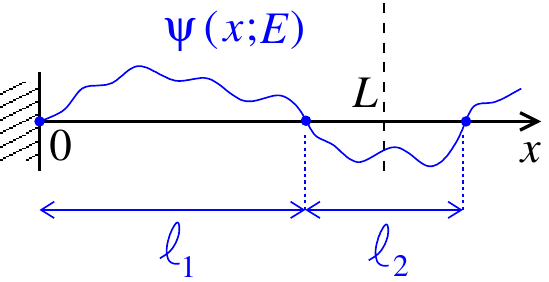}{\scale}{-1cm}
  \caption{\it Solution of the Cauchy problem for different energies, below, equal and above the ground state energy.}
  \label{fig:psi1to3}
\end{figure}

\begin{figure}[!ht]
  \centering
  \includegraphics[width=0.45\textwidth]{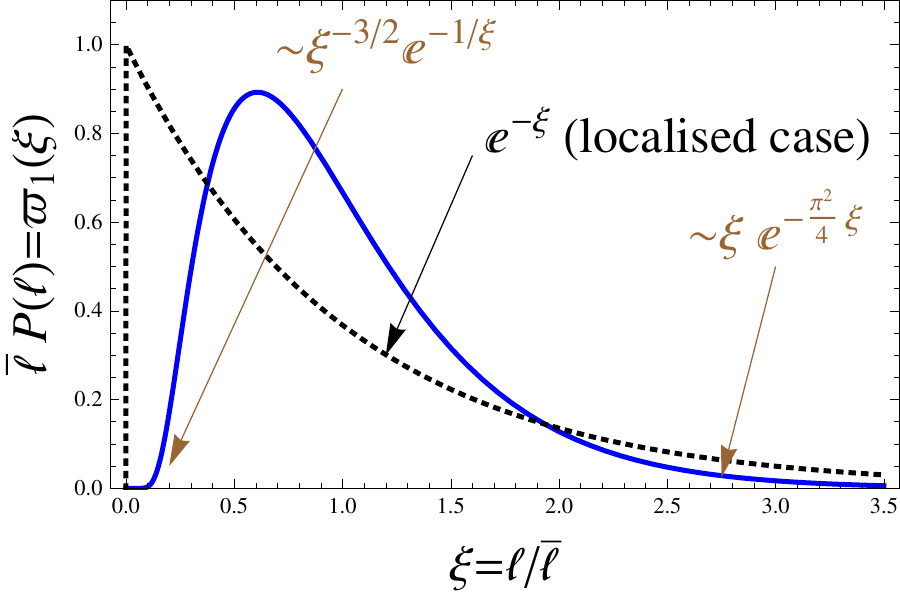}
  \hspace{0.5cm}
  \includegraphics[width=0.45\textwidth]{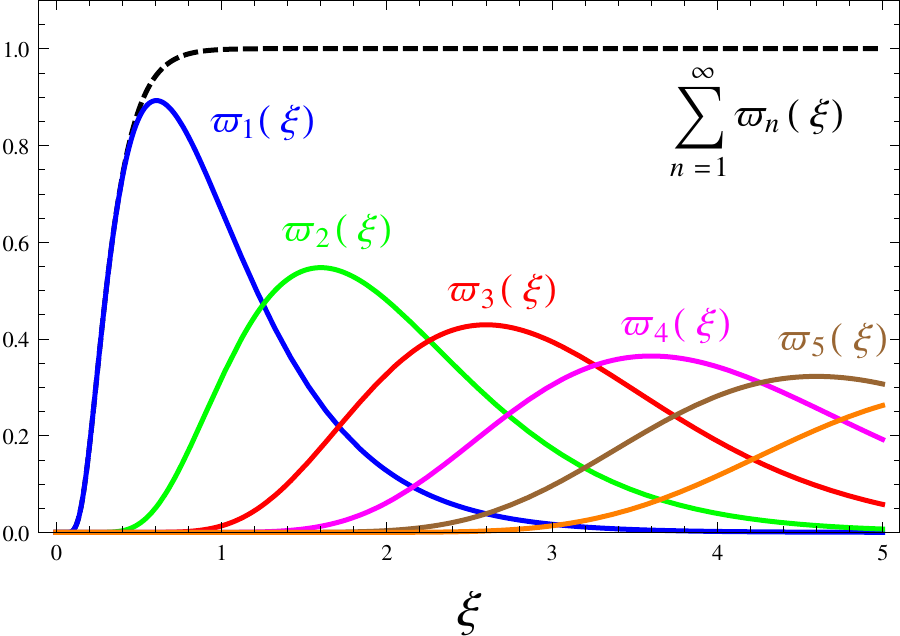}
  \caption{Left~: 
    {\it The distribution $\varpi_1(\xi)$ may be interpreted as the 
    distribution of the rescaled distance $\ell$ between consecutive nodes of the 
    wave function
    $P(\ell)=\varpi_1(\ell/\smean{\ell})/\smean{\ell}$~; the low energy wave function of the SUSY Hamiltonian therefore exhibits repulsion between nodes. For reference, we have also plotted (dotted black) the exponential distribution (absence of repulsion between nodes) characterizing localised eigenstates, the case discussed in \S~2.1 \&~2.2.}
    Right~:
    {\it Distributions $\varpi_n(\xi)$ of the five first rescaled eigenvalues
    $\xi_n=LN_\mathrm{susy}(E_n)$. 
    The sum (dashed line) corresponds to the averaged DoS in a finite interval, after
    spectrum unfolding with respect to bulk DoS~;
    depletion near $\xi=0$ is due to the delocalization.}
    }
  \label{fig:varpin}
\end{figure}

The building brick is now the distribution $P(\ell)$ and we 
explain how it can be calculated.
Introducing the Riccati variable $z=\psi'/\psi-\phi$, we map the
Schr\"odinger equation onto a Langevin like equation
\begin{equation}
  \label{eq:Langevin}
  \deriv{}{x} z(x) = - E - z(x)^2 - 2\,z(x)\,\phi(x)
  \:.
\end{equation}
In this language the distance $\ell_n$ separates two consecutive
divergencies of the
Riccati variable, e.g. $z(0)=+\infty$ and $z(\ell_1)=-\infty$.
The study of $P(\ell)$ is therefore mapped onto a first passage
problem for a stochastic Markovian process, a problem solved by
well-known techniques.
We introduce the auxiliary ``time'' $\mathcal{L}_z$ needed by the
process $z(x)$ in order to reach $-\infty$ for the first time, given
that it has started at $z$. The characteristic function
$h(z;\alpha)=\mean{\EXP{-\alpha\mathcal{L}_z}}$ obeys a backward
Fokker-Planck equation 
\begin{equation}
  \label{eq:BFPE}
  \mathcal{B}_zh(z;\alpha)=\alpha\,h(z;\alpha)
\end{equation}
for boundary conditions $h(-\infty;\alpha)=1$ and
$h'(+\infty;\alpha)=0$, where
$\mathcal{B}_z=2gz\deriv{}{z}z\deriv{}{z}-(E+z^2)\deriv{}{z}$ is the
generator of the diffusion. Finally the distribution $P(\ell)$ can be obtained
by an inverse Laplace transform
\begin{equation}
  \label{eq:ILT}
  h(+\infty;\alpha) = \int_0^\infty\D\ell\, P(\ell)\,\EXP{-\alpha\ell}
  \:.
\end{equation}
An approximation scheme was proposed in Ref.~\cite{Tex00} in order to
solve Eq.~(\ref{eq:BFPE})  in the low energy regime $E\ll g^2$. This 
supposes
that the distributions $W_n(E;L)$ are mostly concentrated on such energy
scales, what is expected to occur when $L\to\infty$.
As a result, it was shown that
\begin{equation}
   h(+\infty;\alpha) \simeq \frac1{\cosh^2\sqrt{\alpha/N_\mathrm{susy}(E)}}
\end{equation}
where $1/\,\mean{\ell\:}=N_\mathrm{susy}(E)\simeq2g/\ln^2(g^2/E)$ is the
IDoSpul of the  
supersymmetric Hamiltonian (\ref{eq:Hsusy}) in bulk ($L=\infty$).
After spectrum unfolding $\xi_n=LN_\mathrm{susy}(E_n)$, the
distributions of the eigenvalues 
$W_n(E;L)=LN'_\mathrm{susy}(E)\,\varpi_n\big(LN_\mathrm{susy}(E)\big)$
are given by  
\begin{equation}
  \label{eq:VarPiN}
  \varpi_n(\xi)  =  \int_\mathscr{B}\frac{\D q}{2\I\pi} \,
  \frac{\EXP{q\xi}}{\cosh^{2n}\sqrt{q}} 
\end{equation}
where $\mathscr{B}$ is a Bromwich contour.
The fact that these distributions deviate from the Poisson
distribution (i.e. $W_n(E;L)$ differ from the generalized Gumbel laws)
signals spectral correlations induced by delocalization.
These distributions were explicitely calculated in
Ref.~\cite{TexHag10} by a generating function method~:
\begin{equation}
  \label{eq:VarPiN2}
  	\varpi_n(\xi)  = 
        \frac{2^{2n}}{\sqrt{\pi}\,\xi^{3/2}}\sum_{m=n}^{\infty}(-1)^{n+m}m
	\left(
	\begin{array}{c}
		 m+n-1\\
		m-n
	\end{array}
	\right)\EXP{-m^2/\xi}
  \:.
\end{equation}
The large $\xi$ behaviour is 
$\varpi_n(\xi)\simeq\frac{\pi^{2n}}{(2n-1)!}\xi^{2n-1}\exp-\frac{\pi^2}{4}\xi$.
The first distributions are represented on Fig.~\ref{fig:varpin}.

The sum of the distributions characterizes the averaged DoS
\begin{equation}
  \label{eq:DoSfi}
  \mean{\varrho(E;L)} = \sum_{n=1}^\infty W_{n}(E;L)
\end{equation}
for a finite interval. It coincides to 
the bulk density of states 
$\mean{\varrho(E;L)}\simeq LN'_\mathrm{susy}(E)$ when eigenstates are
localized on scale smaller than $L$ and do not feel the boundaries. A
log-normal depletion,
$\mean{\varrho(E;L)}\sim\frac1{E\sqrt{gL}}\exp-\frac1{2gL}\ln^2(g^2/E)$, 
is however obtained at low energy 
$E\ll g^2\exp-\sqrt{2gL}$, illustrating 
the sensitivity of low energy delocalized states to the boundaries.
The DoS is represented on Fig.~\ref{fig:varpin} (dashed line)
after spectrum unfolding.
Conversely the boundary sensitivity of the averaged DoS
$\mean{\varrho(E;L)}$ may be used as a localization criterium, what
gives the energy dependence of the localization length
$\xi_E\sim\frac1g\ln^2(g^2/E)$.

Details on the results reviewed in this section can be found in
Refs.~\cite{Tex00,TexHag10}.

\section{Supersymmetry broken by disorder}
\label{sec:breaksusy}

As we have mentioned, the random Hamiltonians $H_\mathrm{scalar}$
and $H_\mathrm{susy}$ present opposite spectral and localization
properties. This observation has led to question the interplay between
the two types of disorder and consider the mixed case
\begin{equation}
  \label{eq:Hmixed}
  H = -\deriv{^2}{x^2} + \phi(x)^2 + \phi'(x) + V(x) 
  \:.
\end{equation}
The case where $\phi$ and $V$ are two Gaussian white noises was
analyzed in Ref.~\cite{HagTex08}, however it turns out that the case
where the potential $V$ describes a random
superposition of repulsive scatterers at random positions,
$V(x)=\sum_i\alpha_i\,\delta(x-x_i)$, can be 
related to an interesting problem in the context of classical
diffusion (discussed in next section). 
In this latter case, the spectral density of the Hamiltonian
(\ref{eq:Hmixed}) may be obtained from the distributions $W_n(E;L)$
obtained in the previous section by using a Lifshits like
argument~\cite{TexHag09}. 
Let us denote by $N(E)$ the IDoSpul of the Hamiltonian
(\ref{eq:Hmixed}). In the limit of large weights $\alpha_i\to\infty$
(in practice $\alpha_i\gg\rho$ \& $g$) the impurities decouple the
intervals free of impurity and impose Dirichlet boundary conditions at
their locations. Thus the spectrum of (\ref{eq:Hmixed}) is given the
addition of spectra of $H_\mathrm{susy}$ on all intervals and we obtain the
DoSpul $N'(E)$ for the mixed disorders by averaging the DoS
(\ref{eq:DoSfi}) of the 
supersymmetric Hamiltonian for a finite length  
\begin{equation}
  \label{eq:LifshitsMultilevel}
  N'(E) \simeq \rho \, 
  \big\langle\, \mean{\varrho(E;L)}\, \big\rangle_L
  \:,
\end{equation}
where $\langle\cdots\rangle_L$ denotes averaging with respect
to the length of interval, namely with an exponential law
$\rho\,\EXP{-\rho L}$ characterizing uncorrelated positions $x_i$.
Using (\ref{eq:VarPiN},\ref{eq:DoSfi}) we get
\begin{equation}
  \label{eq:IDoSmixed}
  N(E) \simeq \frac{\rho }{ \sinh^2\sqrt{\rho/N_\mathrm{susy}(E)} }
  \:.
\end{equation}
Above the threshold energy $E_c=g^2\exp-\sqrt{2g/\rho}$, the IDoS
coincides with the one of the supersymmetric Hamiltonian. Below the
threshold, the Dyson singularity is transformed into a power-law
singularity 
\begin{equation}
  \label{eq:IDoSmixed2}
  N(E) \underset{\rho\to0}{\simeq} 4\rho\, 
  \left(\frac{E}{g^2}\right)^{\sqrt{2\rho/g}}
  \:.
\end{equation}
This power law behaviour has been very well confirmed by numerical
calculations~\cite{TexHag09}. 
Moreover the numerics has
shown that the exponent $\sqrt{2\rho/g}$ seems more general than the
range of application of the Lifshits argument suggests, what we will
comment on in the last section.

\section{Sinai diffusion with weakly concentrated absorbers}
\label{sec:sinaibs}

The study of classical diffusion in a random environment has attracted
a lot of attention in various contexts ranging from  mathematics,
statistical physics (glassy dynamics or polymer physics) and even finance.
It may be described within the framework of Fokker-Planck equation 
$\partial_tP(x;t)=\partial_x(\partial_x-2\phi(x))P(x;t)$, where
$\phi(x)$ is chosen to be  a Gaussian white noise with
$\mean{\phi(x)}=\mu\,g$ and 
$\mean{\phi(x)\phi(x')}-\mean{\phi(x)}\:\mean{\phi(x')}=g\,\delta(x-x')$. 
The model exhibits a rich dynamics as a function of the drift
$\mu$~\cite{BouComGeoLeD90}. In the absence of drift, for $\mu=0$, the
diffusion is controlled by overcoming of potential barriers and is
extremely slow $x(t)\sim\frac1g\ln^2(g^2t)$ (Sinai diffusion~\cite{Sin82}).

The question I would like to discuss now is~: how does the
introduction of absorbing
sites affect the Sinai dynamics~?
Let us go back for a moment to the free diffusion~: in this case the
return probability decays in time as a power law 
$P(x,t|x,0)=1/\sqrt{4\pi t}$. 
The dynamics is slowed down by the introduction of a random force
field, what is reflected in the return probability by an extremely
slow decay $\mean{P(x,t|x,0)}\sim g\ln^{-2}(g^2t)$, related to the
behaviour aforementioned.
On the other hand, when a weak concentration $\rho$ of efficient
absorbers with large absorbing rates $\alpha\gg\rho$ is introduced,
the free power law decay is transformed into an exponential decay 
$\mean{P(x,t|x,0)}\sim\exp-3(\frac{\pi^2}{4}\rho^2t)^{1/3}$ (Lifshits tail).
The question is~: what is the behaviour of the return probability when
both random force field and absorbers are present~? Is the return
probability increased or decreased~?

The answer to this question can be obtained by a simple mapping to the
model analyzed in the previous section. The effect of absorbers on the
Sinai diffusion may be accounted for by adding to the Fokker-Planck
equation a term~:
\begin{equation}
  \label{eq:FPEwA}
  \partial_tP(x;t)=\partial_x \big(\partial_x-2\phi(x) \big) P(x;t) 
  - V(x)\,P(x;t)
\end{equation}
where $V(x)=\sum_i\alpha_i\,\delta(x-x_i)$ describes absorbers located
at a set of positions $\{x_i\}$. The positive coefficient $\alpha_i$
measures the efficiency of the absorber located at $x_i$. 
The Fokker-Planck equation (\ref{eq:FPEwA}) may be mapped onto the
Schr\"odinger equation $-\partial_t\psi(x;t)=H\psi(x;t)$ for the
Hamiltonian (\ref{eq:Hmixed}) thanks to the transformation
$P(x;t)=\psi(x;t)\,\exp\int^x\D{}x'\,\phi(x')$. It follows that the
return probability can be related to the spectral density (DoSpul) 
$\rho(E)=N'(E)$ of
Hamiltonian (\ref{eq:Hmixed}) by 
\begin{equation}
  \label{eq:RelationPDoS}
  \mean{P(x,t|x,0)} = \int_0^\infty\D E\, \rho(E)\,\EXP{-Et}
  \:.
\end{equation}
We immediately deduce the large time behaviour of the return
probability from (\ref{eq:IDoSmixed2})~:
\begin{equation}
  \label{eq:DecayP}
  \mean{P(x,t|x,0)} \sim t^{-\sqrt{2\rho/g}}
  \mbox{ for }
  t\gg t_c
  \:,
\end{equation}
where the scale $t_c=1/E_c=\frac1{g^2}\exp{\sqrt{2g/\rho}}$
 is the time needed by the particle moving in the random force
field to reach the closest absorber $x(t_c)\sim1/\rho$ where 
$x(t)\sim\frac1g\ln^2(g^2t)$.
The power-law can be mostly explained by analyzing the lowest
absorption rate $E_1[\phi(x),L]$ in a finite interval~: the fact that
its distribution presents a log-normal behaviour 
$W_1(E;L)\sim\frac1E\exp-\frac1{2gL}\ln^2(g^2/E)$ 
for $E\to0$ explains the exponent.
The behaviour (\ref{eq:DecayP}) was first obtained in
Ref.~\cite{TexHag09} and later confirmed in Ref.~\cite{LeD09} by
the real space renormalization group method.
 The return probability decays slower than in the free case for a
 sufficiently  weak concentration of absorbers
$\rho<g/8$, whereas it decays faster for $\rho>g/8$. 
Note however that the origin of the decay in the present model is 
mostly due to absorption, as demonstrated in
Ref.~\cite{LeD09} by analyzing the survival probability which was
shown to present the same power law decay up to a logarithmic correction
$\mean{S_{x_0}(t)}=\int\D{}x\,\mean{P(x,t|x_0,0)}\sim (\ln t)\,t^{-\sqrt{2\rho/g}}$.

\section{Concluding remarks}
\label{sec:conclu}

In this brief review, I have analyzed the ordered statistics (extreme
value statistics) for eigenvalues of a random supersymmetric Hamiltonian.
The fact that this quantum Hamiltonian possesses delocalized low energy
eigenstates makes the problem nontrivial
and escape from the universal generalized Gumbel laws obtained for
generic 1D 
random Hamiltonians with a full spectrum of localized states.

These results have found some application in order to analyze the
effect of dilute absorbers on the classical diffusion in a random
environment (Sinai diffusion). We have obtained that the return
probability presents a power law decay, mostly due to absorption. 
This conclusion relies on the
analysis of the spectral density for a mixed random Hamiltonian with a
``supersymmetric part'' $\phi^2+\phi'$, where $\phi(x)$ is a Gaussian
white noise, and a ``scalar part''
$V(x)=\sum_i\alpha_i\,\delta(x-x_i)$. Using a Lifshits like argument,
it was shown that the DoS of this Hamiltonian presents a power law
singularity at low energy $\rho(E)=N'(E)\sim E^{-1+\sqrt{2\rho/g}}$. 
The result is interesting by itself since only few power law spectral
singularities are known in the context of 1D Anderson localization~:
({\it i}) the Halperin singularity, for randomly dropped attractive impurities
of {\it fixed} weights $\alpha_i=-v$ \cite{Hal67,LifGrePas88}~: 
$\rho(E)\sim|E+\frac14v^2|^{-1+2\rho/v}$ for $E\sim-\frac14v^2$. 
({\it ii}) The power law behaviour for the
supersymmetric Hamiltonian $H_\mathrm{susy}$ for a finite
$\mean{\phi(x)}=\mu\,g\neq0$ for which $\rho(E)\sim
E^{-1+\mu}$~\cite{BouComGeoLeD90}.  

Surprisingly, numerical simulations of the model (\ref{eq:Hmixed}) revealed that the exponent 
appearing in the IDoSpul (\ref{eq:IDoSmixed2}) 
is much more robust than one would have expected 
on the basis of the heuristic Lifshits argument.
Examining this observation has led us
to  a solvable version of the model by considering random weights
$\alpha_i$. 
Let us consider the Hamiltonian (\ref{eq:Hmixed}) for a finite drift
$\mean{\phi(x)}=\mu\,g\neq0$ and with exponentially distributed
positive weights $p(\alpha_i)=\frac1v\exp-\frac{\alpha_i}{v}$. 
This model becomes solvable 
for a particular value of the parameter $v=\mean{\alpha_i}$~:
if $v=2g$ the effect of the ``scalar term'' $V(x)$ can be absorbed
in a redefinition of the parameter $\mu$ \cite{TexTou12}
\begin{equation}
  \mu \longrightarrow \nu = \sqrt{\mu^2+\frac{2\rho}{g}}  
\end{equation}
The IDoS of (\ref{eq:Hmixed}) can be obtained by performing this
substitution in the IDoS of $H_\mathrm{susy}$ (that can be found in
Ref.~\cite{BouComGeoLeD90}), therefore~:
\begin{equation}
  N(E) =
  \frac{2g}{\pi^2}\frac1{J_\nu(\sqrt{E}/g)^2+N_\nu(\sqrt{E}/g)^2} 
 \:,
\end{equation}
where $J_\nu(x)$ and $N_\nu(x)$ are Bessel functions~\cite{gragra}.
Using their asymptotic behaviours we obtain the low energy IDoS
$N(E)\simeq\frac{2g}{\Gamma(\nu)^2}\big(\frac{E}{4g^2}\big)^\nu$
for $E\ll g^2$. Therefore
we have recovered the power law with exponent $\sqrt{2\rho/g}$
(setting $\mu=0$) within a solvable model, confirming the robustness of
this exponent valid for arbitrary value of the ratio $\rho/g$.
It is worth emphasizing that the solvable model applies to a regime
with weights $\alpha_i\sim g$ while the analysis of the previous
sections rather relied on the hypothesis that $\alpha_i\gg g,\:\rho$. 
This result also generalizes the analysis to finite drift $\mu\neq0$ (note
that the same exponent $\sqrt{\mu^2+2\rho/g}$ was obtained by
Le~Doussal in 
Ref.~\cite{LeD09} by another method, in a different regime
$\alpha_i\to\infty$).
Details of this analysis will be published in a forthcoming
article~\cite{TexTou12}. 

\section*{Acknowledgements}

The material presented here results from collaborations with Alain Comtet,
Christian Hagendorf and Yves Tourigny to whom I address my deep
thanks. I am grateful to Satya Majumdar for helpful remarks and suggestions.
Nordita Institute at Stokholm is acknowledged for his hospitality
during the program ``{\it Fundations and Applications of non-equilibrium
statistical mechanics}'' (october 2011), 
financially supported by the European Science Fundation.


\appendix

\section{Diffusion constant of a single $d$-dimensional
  Brownian curve}

In Ref.~\cite{TexHag10}, the distribution of the maximal height of a Brownian excursion was
shown to be related to the distributions (\ref{eq:VarPiN},\ref{eq:VarPiN2}),
or more precisely to their sum~$\sum_{n=1}^\infty\varpi_n(\xi)$.

In this appendix, we point out that the distributions 
\begin{equation}
  \label{eq:pin}
 \pi_n(\xi)  =  \int_\mathscr{B}\frac{\D q}{2\I\pi} \,
  \frac{\EXP{q\xi}}{\cosh^{n}\sqrt{q}} 
  \:,  
\end{equation}
generalizing $\varpi_n(\xi)=\pi_{2n}(\xi)$ and considered in Ref.~\cite{TexHag10}, 
are relevant in order to characterize a certain property of the 
$d$-dimensional Brownian motion. 

Let us consider a $d$-dimensional Brownian curve $\vec r(\tau)$ with
$\tau\in[0,t]$ starting from the origin 
$\vec r(0)=\vec0$.
This curve is weighted according to the Wiener measure 
\begin{equation}
  \label{eq:MesureWienerDdim}
  \mathcal{D}\vec r(\tau)\,
  \EXP{-\frac1{4D}\int_0^t\D\tau\, \big(\deriv{\vec r(\tau)}{\tau}\big)^2 }
  \:,
\end{equation}
where $D$ is the diffusion constant, usually defined by 
$D\eqdef\lim_{t\to\infty}\frac{1}{2td}\expect{\vec r(t)^2}$, 
where $\expect{\cdots}$ denotes the expectation with respect to the
Wiener measure (\ref{eq:MesureWienerDdim}).
We now ask the question~: 
is it possible to obtain an estimation of the diffusion constant from
a {\it given realization} of the Brownian path by a time average~?
For this purpose we introduce the \og trajectory-dependent diffusion constant\fg{}
$\mathscr{D}_t\eqdef\frac1{td}\int_0^t\frac{\D\tau}{t}\,\vec r(\tau)\,^2$,
or preferably the dimensionless Brownian functional
\begin{equation}
  \label{eq:5}
  \chi[\vec r(\tau)] \eqdef \frac{\mathscr{D}_t}{D} 
  = \frac1{Dtd} \int_0^t\frac{\D\tau}{t}\,\vec r(\tau)\,^2
  \:.
\end{equation}
Note that this problem was also considered in
Ref.~\cite{BoyDeaMejOsh12}, starting from a different definition of the
trajectory-dependent diffusion constant 
$\int_0^t\frac{\D\tau}{t}\,\frac1{2d\tau}\vec r(\tau)\,^2$.

The characteristic function of the functional $\chi[\vec r\,]$ may be
written thanks to a path integral
\begin{equation}
  \label{eq:2}
\Gamma(p) \eqdef \expect{ \EXP{-p\,\chi[\vec r(\tau)]} } 
  =  \int\D\vec R   \int_{\vec r(0)=\vec0}^{\vec r(t)=\vec R}
   \hspace{-0.5cm} \mathcal{D}\vec r(\tau)\,
  \EXP{
     -\int_0^t\D\tau\,
     \big[
     \frac1{4D}\dot{\vec r}(\tau)^2
      + \frac{p}{Dt^2d}\vec r(\tau)\,^2
     \big]
      }
\end{equation}
We recognize the integral of the imaginary time propagator 
$\bra{\vec R}\EXP{-tH}\ket{\vec0}$ for a quantum harmonic oscillator
with mass $m\to1/(2D)$ and pulsation
$\omega\to\frac2t\sqrt{\frac{p}{d}}$.
Using its well-known expression~\cite{FeyHib65} we straightforwardly obtain
\begin{equation}
   \Gamma(p) = \frac1{\cosh^{\frac d2}\left(2\sqrt{p/d}\right)}
  \:.
\end{equation}
We check that $\expect{\chi[\vec r\,]}=1$ as it should. The variance is given by 
$\mathrm{Var}(\chi[\vec r\,])=\frac{4}{3d}$ and vanishes in the limit of large
dimension, as expected from the central limit theorem.
In the case of even dimensions, we recognize the Laplace transform of the
distribution (\ref{eq:pin}).
Therefore we deduce that the distribution of the functional 
$\chi[\vec r\,]$ is 
\begin{equation}
  P_d( \chi ) = \frac d4\: \pi_{\frac d2}\!\left( \frac d4\chi \right) 
  \hspace{0.5cm}\mbox{for dimension } d \mbox{ even}
  \:.   
\end{equation}
Expressions of the distributions $\pi_n(\xi)$ can be found in Ref.~\cite{TexHag10}.

\section*{References}



\end{document}